\documentclass[twocolumn,showkeys,aps,prb,showpacs]{revtex4-1}
\usepackage{graphicx}
\usepackage{amsmath}
\usepackage[CJKbookmarks,dvipdfm,colorlinks,linkcolor=blue,citecolor=blue]{hyperref}

\begin{document}
\title{Biaxial tensile strain tuned  up-and-down behavior on lattice thermal conductivity in $\beta$-AsP monolayer}
\author{San-Dong Guo}
\affiliation{School of Physics, China University of Mining and
Technology, Xuzhou 221116, Jiangsu, China}

\begin{abstract}
Various two-dimensional (2D)  materials with graphene-like buckled structure emerge,  and the $\beta$-phase AsP monolayer  has been recently proposed to be thermodynamically stable from first-principles calculations.  The studies of thermal transport are very useful for these 2D materials-based nano-electronics devices. Motivated by this, a comparative study of strain-dependent phonon transport of  AsP monolayer is performed by solving the linearized phonon Boltzmann equation within the single-mode relaxation time approximation (RTA).  It is found that the lattice thermal conductivity ($\kappa_L$) of AsP monolayer is very close to one of As monolayer with similar buckled structure, which is due to neutralization between  the reduce of phonon lifetimes  and group velocity enhancement from As to AsP monolayer. The corresponding room-temperature sheet thermal conductance of AsP monolayer is  152.5 $\mathrm{W K^{-1}}$.
It is noted that the increasing tensile strain  can harden  long wavelength out-of-plane  (ZA) acoustic mode, and soften the in-plane longitudinal acoustic (LA) and transversal acoustic (TA) modes. Calculated results show that $\kappa_L$ of AsP monolayer  presents a  nonmonotonic up-and-down behavior with increased strain. The  unusual strain dependence is due to the competition among  reduce  of phonon group velocities,   improved phonon lifetimes of ZA mode and nonmonotonic up-and-down  phonon lifetimes of TA/LA mode. It is found that acoustic branches dominate the $\kappa_L$ in considered  strain range, and the contribution from ZA branch increases with increased strain, while it is opposite for TA/LA branch.
By analyzing cumulative $\kappa_L$  with respect to phonon mean free path (MFP), tensile  strain can  modulate effectively size effects on $\kappa_L$ in AsP monolayer. Our work  enriches the studies of thermal transports of 2D  materials with graphene-like buckled structure, and strengthens  the idea to engineer thermal transport properties  by simple mechanical strain, and stimulates  further experimental works to synthesize AsP monolayer.

\end{abstract}

\pacs{72.15.Jf, 71.20.-b, 71.70.Ej, 79.10.-n ~~~~~~~~~~~~~~~~~~~~~~~~~~~~~~~~~~~Email:guosd@cumt.edu.cn}

\maketitle

\section{Introduction}
Due to great importance for new-generation high performance electronic devices, transition-metal dichalcogenide (TMD)\cite{q7}, group IV-VI\cite{q8}, group-VA\cite{q9,q10}  and group-IV\cite{q11}  monolayers have been widely investigated both theoretically and experimentally.
Compared with graphene as representative 2D material, group-VA monolayers are semiconductors with significant intrinsic band gaps, which have potential applications for  electronic, optoelectronic, thermoelectric and energy devices\cite{t1}.
 Phosphorene (P monolayer) is a classic group-VA monolayer with  a direct band gap of 2.0 eV and a high hole mobility above $10^4$ $\mathrm{cm^2~V^{-1} ~s^{-1}}$\cite{t2,t3,t4,t5}. Shortly after phosphorene, group-VA arsenene (As monolayer), antimonene (Sb monolayer) and bismuthene (Bi monolayer) with desirable stability and high carrier mobility are predicted  by the first-principle calculations\cite{q9}. Subsequently, antimonene of them  is  fabricated
experimentally on various substrates via van der Waals epitaxy growth\cite{t8,q10,q1000}. Recently, AsSb, SbBi and AsP monolayers  have also been identified as novel 2D semiconductors in theory\cite{t9,t10,t11,t12}. Under biaxial  strain, quantum spin Hall insulator can be achieved  in
monolayer $\beta$-BiSb and $\beta$-SbAs\cite{t9,t11}. It is also predicted that  giant tunable Rashba spin splitting can be realized in  BiSb monolayer\cite{t10}. The AsP monolayer with direct band gap and very high  mobility is predicted as a promising 2D solar cell donor\cite{t12}.

\begin{figure}
  \includegraphics[width=5.5cm]{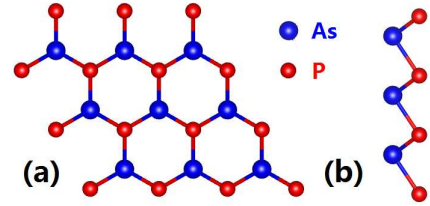}
  \caption{(Color online) Top (Left) and side (Right) views of the crystal structure of AsP monolayer.}\label{st}
\end{figure}

\begin{figure*}
  \includegraphics[width=12cm]{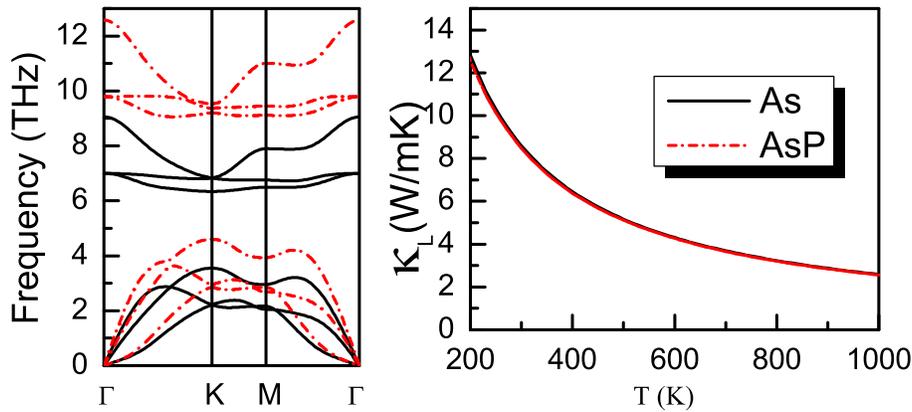}
  \caption{(Color online) Phonon band structures and lattice thermal conductivities of As and AsP monolayers.}\label{ph}
\end{figure*}

\begin{figure}
  \includegraphics[width=8cm]{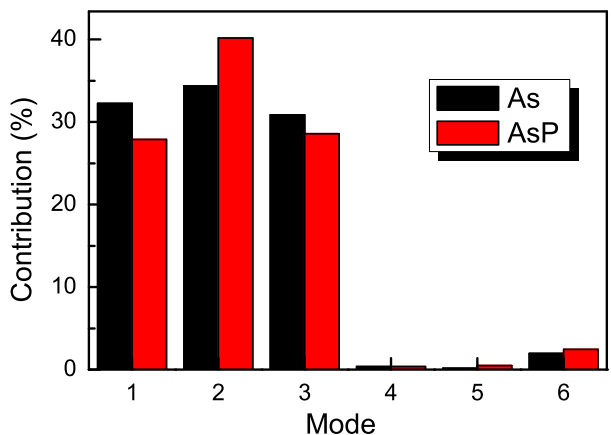}
  \caption{(Color online) The phonon modes contributions toward total lattice thermal conductivity of As and AsP monolayers; The  1, 2, 3 represent  ZA, TA and LA acoustic phonon branches, and 4, 5, 6 for optical branches.}\label{mode}
\end{figure}

As is well known,  thermal transport is  a key
factor for  the  performance of  nano-devices.  To effectively remove the accumulated heat,  a high thermal conductivity is needed.  However,  a  material with low thermal conductivity has potential application in thermoelectric field.
In theory, thermal transports  of group-VA monolayers   have  been widely studied for both  electron and phonon parts\cite{l1,l2,l3,l3-1,l3-2,l3-3,l3-4,l3-5,l3-6}. The  $\alpha$-arsenene and  antimonene show  anisotropic $\kappa_L$  along the zigzag and armchair directions\cite{l2,l3-1}. It has been proved  that chemical functionalization  is a effective way to reduce $\kappa_L$ of antimonene\cite{l3-3}.
The thermoelectric properties of group-VA  monolayers have also been performed by the first principle calculations\cite{l3-2,l3-4,l3-6}, and suggest they are potential candidates for thermoelectric application.  Compared with many 2D materials, bismuthene has very low $\kappa_L$, which is very important for thermoelectric application\cite{l3-2}. It is also predicted that  SbAs monolayer has lower $\kappa_L$  than arsenene or  antimonene\cite{l3-5}.
In  nanoscale devices,  the  residual strain usually exists in real applications\cite{l111}, which can produce important effects on the intrinsic physical properties of 2D materials. The strain dependent transport properties have been investigated in lots of 2D materials\cite{l3,l6,l7,l8,l9,l10,l11,l12,l13}. The power factor can be improved by strain in $\mathrm{MoS_2}$,  $\mathrm{PtSe_2}$ and $\mathrm{ZrS_2}$ monolayers due to bands converge\cite{l6,l7,l8}.
Tensile strain can induce strong size effects on $\kappa_L$ of antimonene, silicene, germanene and stanene\cite{l3,l11}. With  increased  strain, the $\kappa_L$ shows monotonous decrease (penta-graphene),   up-and-down (penta-$\mathrm{SiC_2}$) and jump (penta-$\mathrm{SiN_2}$) behavior with similar penta-structures\cite{l10}, which is due to the competition between the change of
phonon group velocities and phonon lifetimes of acoustic phonon branches, along with  the unique structure transition
for penta-$\mathrm{SiN_2}$. The $\kappa_L$ increases with strain increasing  ($<$6\%)  in  antimonene\cite{l3}, while decreases in  $\mathrm{MoTe_2}$\cite{l13}.

In this work, the strain-dependent phonon transport in AsP monolayer is performed  from a combination of  first-principles calculations and the linearized phonon Boltzmann equation.  The $\kappa_L$ of AsP monolayer approximates  to one of As monolayer  due to neutralization between  the reduce of phonon lifetimes  and group velocity enhancement from As to AsP monolayer. As the strain increases, the $\kappa_L$ of AsP monolayer  shows a  nonmonotonic up-and-down behavior, which is due to the competition among  reduce  of phonon group velocities,   enhanced  phonon lifetimes of ZA branch and nonmonotonic up-and-down  phonon lifetimes of TA/LA  branch.

\begin{figure}
  \includegraphics[width=7.8cm]{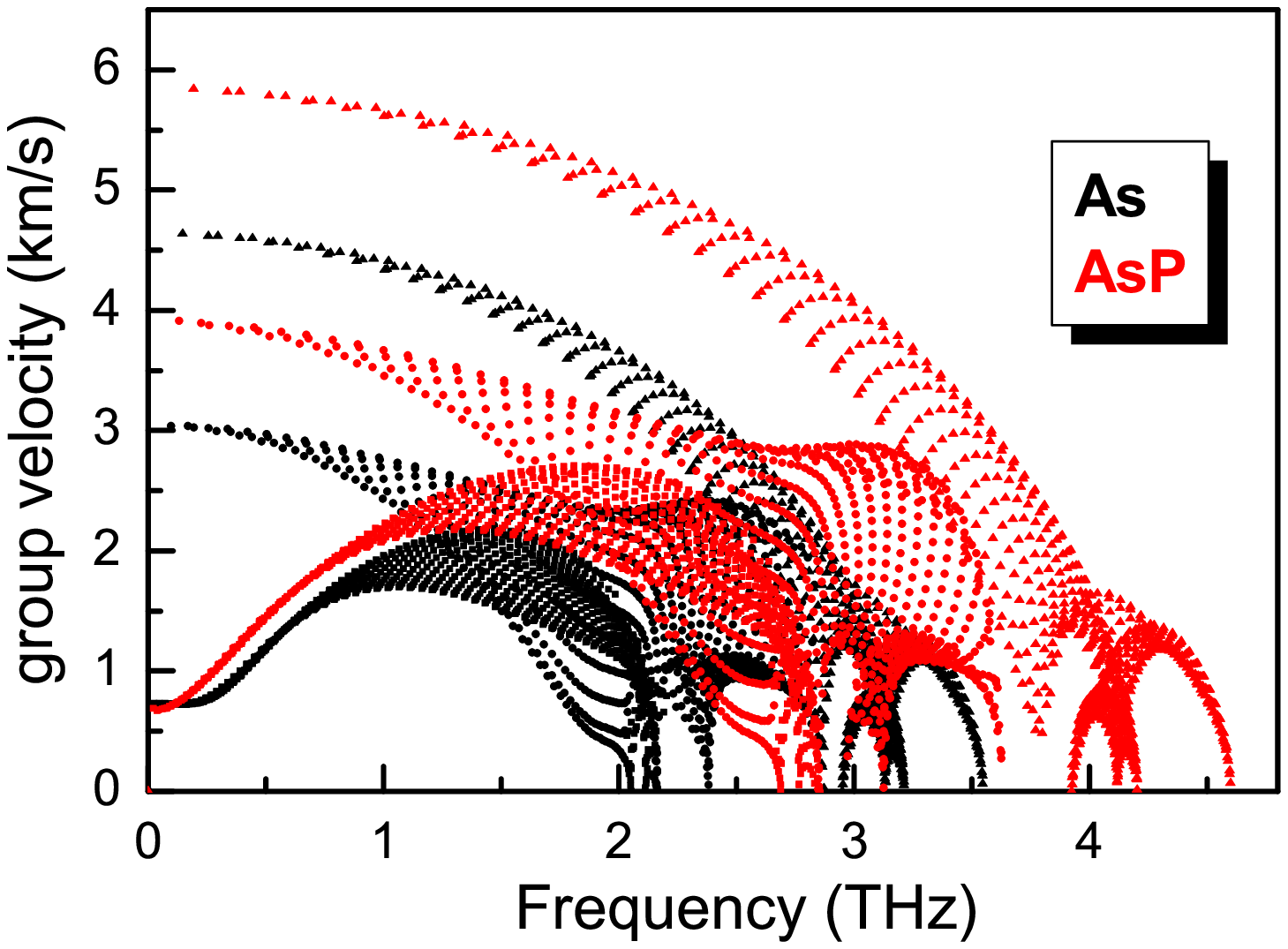}
  \includegraphics[width=8cm]{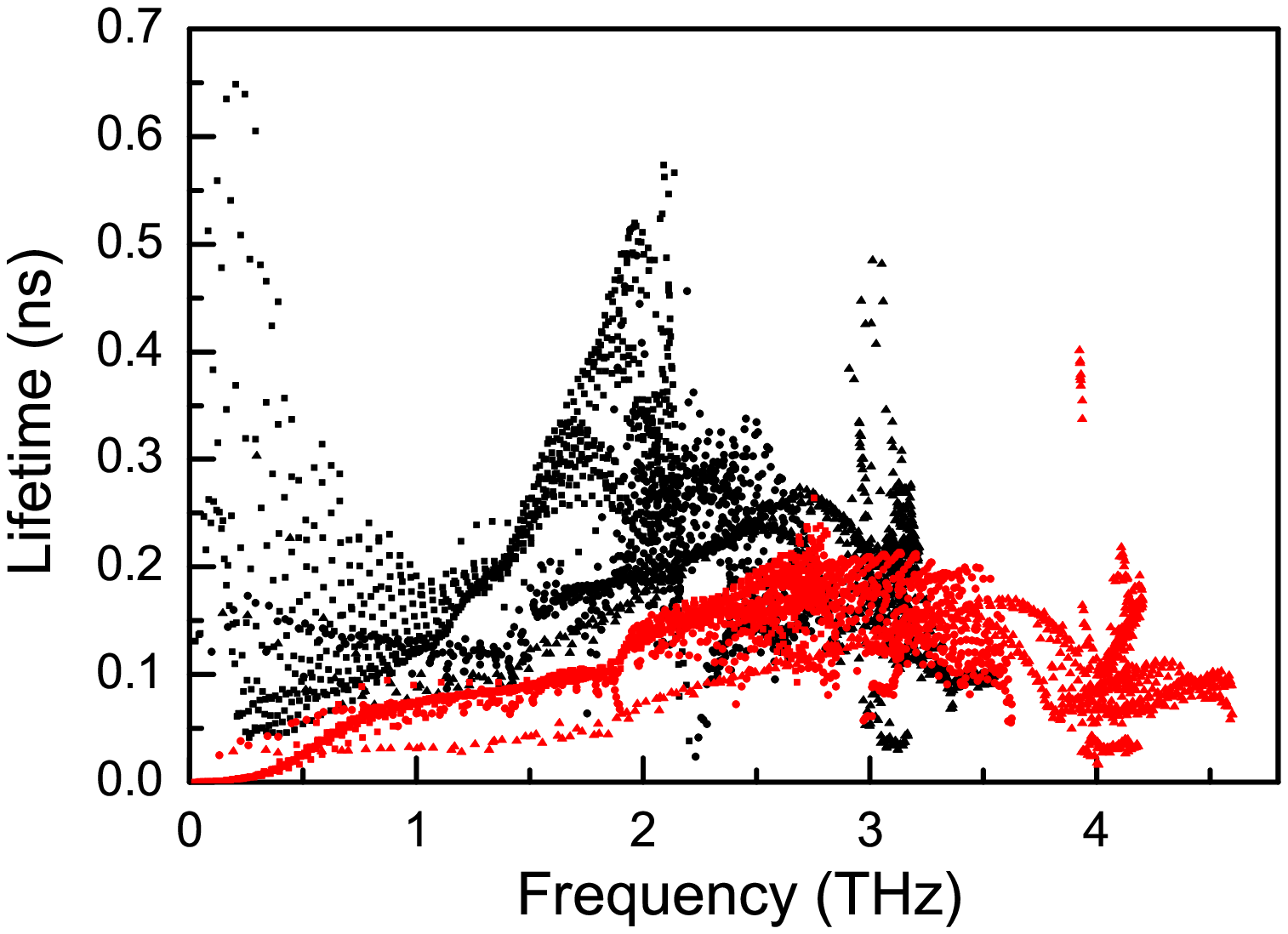}
  \caption{(Color online) The phonon mode group velocities  and phonon lifetimes of As (Black) and AsP (Red)  monolayers in the first BZ  for ZA (square symbol), TA (circle symbol) and LA (UpTriangle symbol) acoustic branches.}\label{vt}
\end{figure}

The rest of the paper is organized as follows. In the next
section, we shall give our computational details about  phonon transport calculations. In the third section, we shall present phonon transports of As and AsP monolayers, and  strain effect on $\kappa_L$ of AsP monolayer. Finally, we shall give our  conclusions in the fourth section.

\section{Computational detail}
  The first-principles calculations are performed within projector augmented-wave method, as implemented in the VASP code\cite{pv1,pv2,pv3}.
The  generalized gradient approximation   of Perdew-Burke-Ernzerhof (PBE-GGA) is adopted  as exchange-correlation functional\cite{pbe}.
  A 20 $\times$ 20 $\times$ 1 k-mesh is used during structural relaxation,  with  a Hellman-Feynman force convergence criteria $10^{-4}$ eV/ $\mathrm{{\AA}}$. A plane-wave basis set is employed, and  the
kinetic energy cutoff is set as  450 eV, and  the electronic stopping criterion is $10^{-8}$ eV.
The 3s and 3p electrons of P,  and  4s and 4p electrons of As are treated as valance ones.
The  lattice thermal conductivity is carried out
by using Phono3py+VASP codes\cite{pv1,pv2,pv3,pv4}.
By solving linearized phonon Boltzmann equation, the $\kappa_L$ is computed with single-mode RTA by  the Phono3py code\cite{pv4}. The $\kappa_L$ can be expressed as:
\begin{equation}\label{eq0}
    \kappa=\frac{1}{NV_0}\sum_\lambda \kappa_\lambda=\frac{1}{NV_0}\sum_\lambda C_\lambda \nu_\lambda \otimes \nu_\lambda \tau_\lambda
\end{equation}
in which  $\lambda$, $N$ and  $V_0$ are  phonon mode, the total number of q points sampling Brillouin zone (BZ) and  the volume of a unit cell, and  $C_\lambda$,  $ \nu_\lambda$, $\tau_\lambda$   is the specific heat,  phonon velocity,  phonon lifetime.
The phonon lifetime $\tau_\lambda$ can be attained  by  phonon linewidth $2\Gamma_\lambda(\omega_\lambda)$ of the phonon mode
$\lambda$:
\begin{equation}\label{eq0}
    \tau_\lambda=\frac{1}{2\Gamma_\lambda(\omega_\lambda)}
\end{equation}
The $\Gamma_\lambda(\omega)$  takes the form analogous to the Fermi golden rule:
\begin{equation}
\begin{split}
   \Gamma_\lambda(\omega)=\frac{18\pi}{\hbar^2}\sum_{\lambda^{'}\lambda^{''}}|\Phi_{-\lambda\lambda^{'}\lambda^{''}}|^2
   [(f_\lambda^{'}+f_\lambda^{''}+1)\delta(\omega
    -\omega_\lambda^{'}-\\\omega_\lambda^{''})
   +(f_\lambda^{'}-f_\lambda^{''})[\delta(\omega
    +\omega_\lambda^{'}-\omega_\lambda^{''})-\delta(\omega
    -\omega_\lambda^{'}+\omega_\lambda^{''})]]
\end{split}
\end{equation}
in which $f_\lambda$  and $\Phi_{-\lambda\lambda^{'}\lambda^{''}}$ are the phonon equilibrium occupancy and the strength of interaction among the three phonons $\lambda$, $\lambda^{'}$, and $\lambda^{''}$ involved in the scattering.
Based on the supercell approach  with finite atomic displacement
of 0.03 $\mathrm{{\AA}}$,  the second-order interatomic force constants (IFCs) can be attained
by using   a 5 $\times$ 5 $\times$ 1   supercell   with k-point meshes of 2 $\times$ 2 $\times$ 1.
According to second-order  harmonic IFCs, phonon dispersions can be calculated by  Phonopy package\cite{pv5}, determining the allowed three-phonon scattering processes. The third-order  IFCs can be attained
by using   a 4 $\times$ 4 $\times$ 1  supercell   with k-point meshes of 3 $\times$ 3 $\times$ 1, which is related to the three-phonon scattering rate.
To compute accurately lattice thermal conductivity,  the reciprocal spaces of the primitive cells  are sampled by 100 $\times$ 100 $\times$ 1 meshes .

For 2D material, the calculated  $\kappa_L$  depends on the length of unit cell along z direction\cite{2dl}.  They  should be normalized by multiplying $Lz/d$, where  $Lz$ is the length of unit cell along z direction,  and $d$ is  the thickness of 2D material.  However, the $d$  is not well defined,  for example graphene.  In this work, the $Lz$=18 $\mathrm{{\AA}}$  is used as $d$. By $\kappa$ $\times$ $d$,  the thermal sheet conductance can be attained.

\begin{figure}
  \includegraphics[width=8cm]{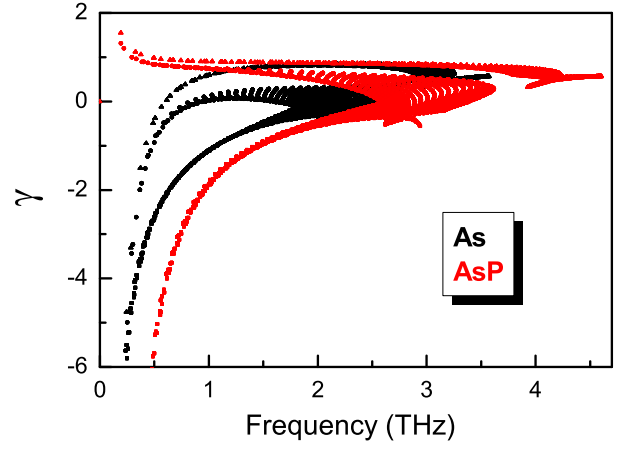}
  \caption{(Color online) The phonon mode group Gr$\ddot{u}$neisen parameter
($\gamma$) of As (Black) and AsP (Red)  monolayers in the first BZ  for ZA (square symbol), TA (circle symbol) and LA (UpTriangle symbol) acoustic branches.}\label{r}
\end{figure}

\section{MAIN CALCULATED RESULTS AND ANALYSIS}
Like As monolayer, the  $\beta$-AsP monolayer (No.156) has a graphene-like buckled honeycomb structure, which  has
 lower symmetry compared with As  monolayer (No.164). The  AsP monolayer can be built  by replacing  one sublayer of As monolayer with P atoms.
The schematic crystal structure of AsP monolayer is plotted in \autoref{st}, and the unit cell  is constructed with the vacuum region of more than 15 $\mathrm{{\AA}}$ to avoid spurious interaction between neighboring layers.
The optimized lattice constant (buckling parameter) is $a$=$b$=3.45 $\mathrm{{\AA}}$   ($h$=1.33 $\mathrm{{\AA}}$),  which is about 4.4\% (5.0\%) lower than that of  As monolayer.  The elastic properties are also studied, and  two independent elastic
constants $C_{11}$=$C_{22}$ and $C_{12}$ are  calculated using the stress-strain relations. The 2D Young¡¯s moduli $Y^{2D}$   and shear modulus $G^{2D}$ are given\cite{ela}:
\begin{equation}\label{e1}
Y^{2D}=\frac{C_{11}^2-C_{12}^2}{C_{11}}
\end{equation}
\begin{equation}\label{e1}
G^{2D}=C_{66}
\end{equation}
\begin{equation}\label{e1}
C_{66}=(C_{11}-C_{12})/2
\end{equation}
The corresponding Poisson's ratios can be expressed as:
\begin{equation}\label{e1}
\nu^{2D}=\frac{C_{12}}{C_{11}}
\end{equation}
The  related data  of As and AsP  monolayers are listed in \autoref{tab1}, which agree well with available  theoretical values\cite{q9,t12}.
It is found that $C_{11}$, $G^{2D}$ and $Y^{2D}$ of AsP monolayer are very close to ones of As monolayer, while $C_{12}$ and $\nu$ are lower  for AsP than As monolayer.  The $G^{2D}$ and $Y^{2D}$ of AsP monolayer
are significantly lower than ones of h-BN and graphene\cite{ela}, which is due to weaker As-P bonds  compared to B-N and C-C bonds. However, those of AsP monolayer are larger than ones of BiSb monolayer with similar crystal structure\cite{t10}.
\begin{table*}
\centering \caption{For As and AsP monolayers,  the lattice constant $a$ ($\mathrm{{\AA}}$) and  the buckling parameter $h$ ($\mathrm{{\AA}}$); the elastic constants $C_{ij}$, shear modulus $G^{2D}$,  Young's modulus $Y^{2D}$ in $\mathrm{Nm^{-1}}$, and Poisson's ratio $\nu$
dimensionless. }\label{tab1}
  \begin{tabular*}{0.96\textwidth}{@{\extracolsep{\fill}}cccccccccc}
  \hline\hline
Name& $a$ &$h$& $C_{11}=C_{22}$ &  $C_{12}$& $C_{66}=G^{2D}$&$Y^{2D}$& $\nu$\\\hline\hline
$\mathrm{As}$& 3.61 & 1.40 &57.24  &18.90   &19.17&51.00& 0.33\\\hline
$\mathrm{AsP}$&3.45& 1.33&55.08&13.12 &20.98&51.95&0.24\\\hline\hline
\end{tabular*}
\end{table*}
\begin{figure}
  \includegraphics[width=8cm]{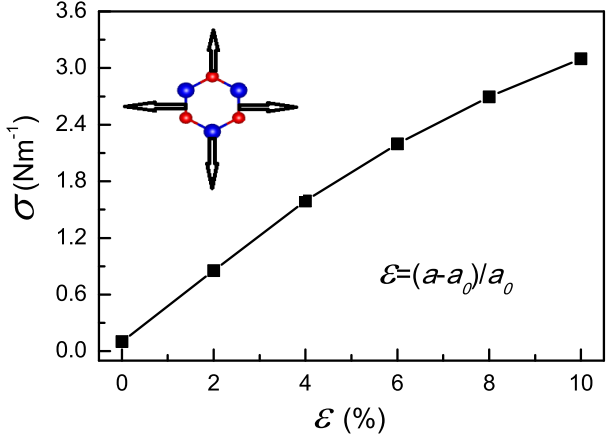}
  \caption{(Color online) Stress vs strain for  AsP monolayer  with strain from  0\% to 10\%.}\label{stress}
\end{figure}
 \begin{figure}
  \includegraphics[width=8cm]{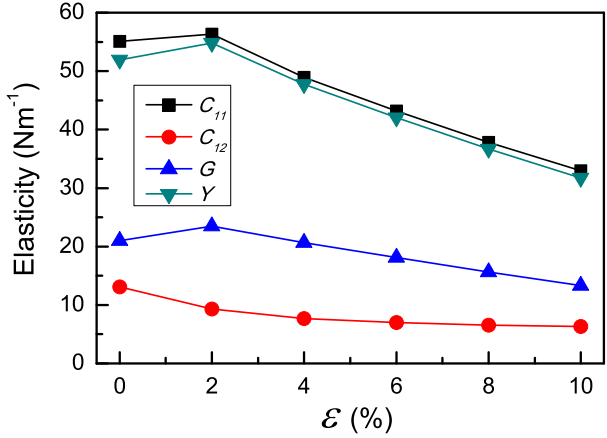}
  \caption{(Color online) The elastic constants $C_{ij}$, Young¡¯s moduli $Y$ and shear modulus $G$  vs strain for  AsP monolayer with strain from  0\% to 10\%.}\label{elastic}
\end{figure}

The phonon dispersion curves of As and AsP monolayers along  high symmetry path ($\Gamma$-K-M-$\Gamma$)   are shown   in \autoref{ph}, which agree well with
previous theoretical results\cite{q9,t12}.The  3 acoustic
and 3 optical phonon branches can be observed due to two atoms in unit cell. It is the general feature of 2D material that  LA and TA branches are linear near the $\Gamma$ point, while ZA branch  deviates from linearity\cite{l3,l6,l7,l8,l9,l10,l11,l12,l13}.
It is clearly seen that there is an acoustic and optical (ao) phonon band gap for both As and AsP monolayer. The ao gap  of AsP monolayer is as large as 4.46 THz, which is very close to the width of acoustic branch (4.60 THz), while the
gap is  2.79 THz for As monolayer, smaller than the range
of acoustic branch  (3.55 THz).  It is noted that the ao gap of As monolayer is  due to the violation of  reflection symmetry selection
rule in the harmonic approximation\cite{m0}, while the gap of AsP monolayer  is due to  mass differences between the constituent atoms besides  the violation of  reflection symmetry selection\cite{m1,m2}. Compared with As monolayer,  acoustic modes of AsP monolayer   become stiff, which mean improved group velocities.
The optical branches of AsP monolayer overall move toward higher energy with respect to ones of As monolayer, which results in larger ao gap for AsP than As monolayer.

\begin{figure*}
  \includegraphics[width=12cm]{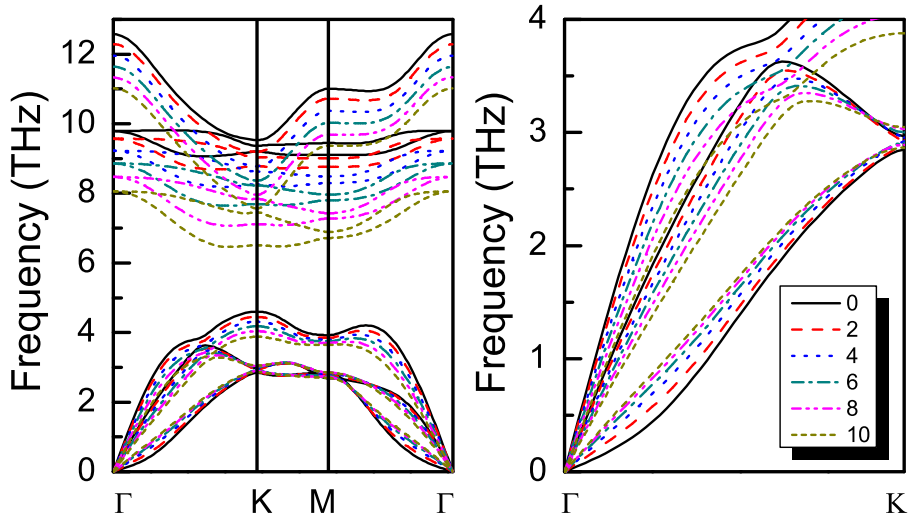}
  \caption{(Color online) Left: phonon dispersion curves of AsP monolayer with strain from 0\% to 10\%. Right: the  acoustic phonon branches  along the $\Gamma$-K direction.}\label{ph-c}
\end{figure*}

 Within  RTA, the lattice  thermal conductivities of  As and AsP monolayers as a function of temperature are plotted in \autoref{ph}. It is found that the $\kappa_L$ of AsP monolayer  is  slightly lower than that of As monolayer, and the room-temperature $\kappa_L$ is 8.47 $\mathrm{W m^{-1} K^{-1}}$ and 8.59 $\mathrm{W m^{-1} K^{-1}}$  with the same thickness $d$ (18 $\mathrm{{\AA}}$). Their  thermal sheet conductance\cite{2dl} is  152.5 $\mathrm{W K^{-1}}$  and  154.6 $\mathrm{W K^{-1}}$, respectively. Our calculated $\kappa_L$  of As monolayer is  in the range of previous ones\cite{l2,m0}.
 It is expected that the $\kappa_L$ of AsP monolayer is higher than one of SbAs monolayer (28.8  $\mathrm{W K^{-1}}$)\cite{l3-5}.
 The relative contributions of every phonon  mode to the total $\kappa_L$ for  As and AsP monolayers (300 K) are plotted in \autoref{mode}.
 It is clearly seen that the acoustic branches of both As and AsP monolayers  dominate $\kappa_L$, about 97.5\%  and  96.7\%.
For As monolayer, the contribution among  three acoustic branches  has little difference, about 32.3\%, 34.4\% and 30.9\% from ZA to TA to LA branch.
For AsP monolayer, the TA branch (40.2\%) has larger contribution than ZA (27.9\%) and LA (28.6\%) branches.
  This is  obviously different from SbAs monolayer with very little contribution  to the total $\kappa_L$ from ZA branch, only 2.4\%\cite{l3-5}.

To understand the almost the same $\kappa_L$ beween As and AsP monolayer,  we only show  acoustic phonon mode group velocities  and  lifetimes   in \autoref{vt} due to dominant contribution from acoustic branches.  It is clearly seen that most of group velocities for AsP monolayer are larger than those of As monolayer
due to stiffened acoustic phonon modes, which results in the increase  of $\kappa_L$.
From As  to AsP monolayer, the largest phonon group velocity  of  LA/TA modes increases  from 3.04/4.63 $\mathrm{km s^{-1}}$ to
3.91/5.84 $\mathrm{km s^{-1}}$  near the $\Gamma$ point, and  the  largest phonon group velocity of ZA branch changes from 2.18 $\mathrm{km s^{-1}}$ to  2.71 $\mathrm{km s^{-1}}$. However, most of phonon lifetimes of  AsP monolayer are shorter than those of As monolayer, which results in the decrease   of $\kappa_L$.
The shorter phonon lifetimes for AsP  than As monolayer can be understood by phonon anharmonicity. The anharmonic nature of structures can be roughly quantified by the Gr$\ddot{u}$neisen parameter ($\gamma$), and phonon mode Gr$\ddot{u}$neisen parameters of As and AsP monolayer  are plotted in \autoref{r}.
The magnitude of $\gamma$ for AsP monolayer is obviously larger
than As monolayer, which means stronger phonon anharmonicity in AsP monolayer, giving rise to the shorter phonon lifetime of AsP monolayer. In a word,
the reduce of phonon lifetimes from As to AsP monolayer  neutralizes the positive
effect of the group velocity enhancement, producing almost the same $\kappa_L$ beween As and AsP monolayer.
\begin{figure}
  \includegraphics[width=8cm]{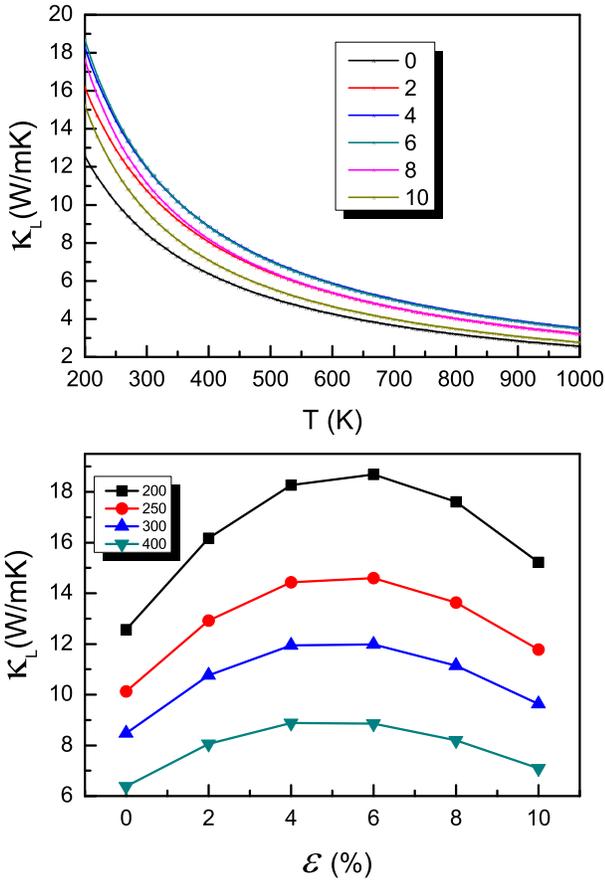}
  \caption{(Color online) Top: the lattice thermal conductivities  of AsP monolayer with strain from  0\% to 10\% as a function of temperature. Bottom: the lattice thermal conductivities (200, 250, 300 and 400 K)  of AsP monolayer  versus  strain.}\label{kl-c}
\end{figure}

Strain dependence of $\kappa_L$ has been investigated in many 2D materials\cite{l3,l6,l7,l8,l9,l10,l11,l12,l13}. Here, the $\varepsilon=(a-a_0)/a_0$ is defined to describe biaxial strain effects on $\kappa_L$ of AsP monolayer, where $a_0$ is the unstrain lattice constant. The calculations are  carried out   on  six values of $\varepsilon$ with tensile strain  from 0\% to 10\%. Firstly, the stress is calculated  as a function of strain, which   is shown in \autoref{stress}. In considered strain range,  the stress is relatively small, about  3.1 $\mathrm{Nm^{-1}}$ for 10\% strain, which  can be easily realized experimentally.
The relatively small stress caused by tensile strain is also found in SbAs monolayer\cite{l3-5}.
The elastic constants $C_{ij}$, Young¡¯s moduli $Y$ and shear modulus $G$  vs strain for  AsP monolayer with strain from  0\% to 10\% are plotted in \autoref{elastic}.  It is found that $C_{11}$, $Y$ and $G$ firstly increase with strain from 0\% to 2\%, and then decrease. However, the $C_{12}$  monotonically decreases from 0\% to 10\% strain. In considered strain range, they all satisfy the  Born  criteria of mechanical stability\cite{ela}:
\begin{equation}\label{e1}
C_{11}>0,~~ C_{66}>0
\end{equation}

The phonon dispersions along  high symmetry directions with strain from 0\% to 10\% are plotted \autoref{ph-c}. Calculated results show that no imaginary frequencies  are observed with strain from 0\% to 10\%, which indicates
that strained AsP monolayer is dynamically stable.
It is clearly seen that the dispersions of both TA
and LA branches are softened with increasing strain, resulting in
the reduction of phonon group velocity.  However, the  dispersion  of ZA branch  near $\Gamma$ point is stiffened with strain increasing,  indicating the
enhancement of phonon group velocity. Similar change of acoustic group velocities can be found in penta-graphene\cite{l10} and $\mathrm{MoTe_2}$\cite{l13}.
It is noted that the
quadratic nature of  ZA mode near the $\Gamma$ point turns into a straight line. Similar phenomenon can be observed in $\mathrm{MoTe_2}$\cite{l13}.
The  dispersions of optical branches overall move toward low energy  with increasing  strain from 0\% to 10\%, which can be explained by   less strongly interacting between atoms. The ao gap decreases from 4.46 THz to 2.58 THz with strain from 0\% to 10\%,  which has important effects on acoustic+acoustic$\rightarrow$optical (aao) scattering

\begin{figure}
  \includegraphics[width=8cm]{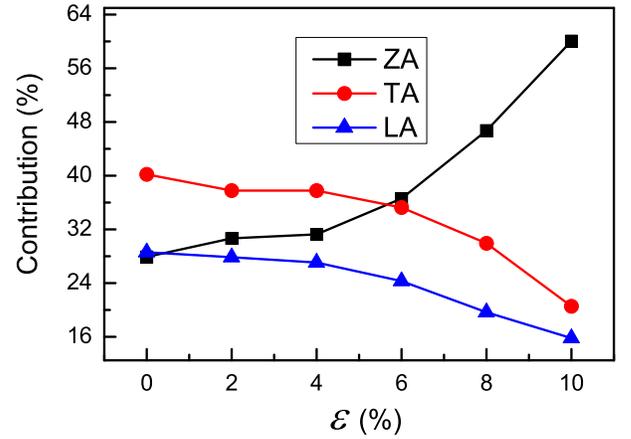}
  \caption{(Color online)The phonon modes contributions of ZA, TA and LA branches to total lattice thermal conductivity as a function of strain.}\label{mode-c}
\end{figure}

The $\kappa_L$ of  AsP monolayer in the strain range of 0\% - 10\% as a function of temperature are shown in \autoref{kl-c}, along with  $\kappa_L$ vs strain at the temperature of  200, 250, 300 and 400  K. In considered strain range, $\kappa_L$ firstly increases with strain increasing, and then decreases. The critical strain is about 5\%. Similar up-and-down behavior is also found in penta-$\mathrm{SiC_2}$\cite{l10} and bilayer graphene\cite{bg}, and their  peak value occurs at 5\% and 3\% tensile strain. The contribution of each mode to total $\kappa_L$ is calculated at different strain, and the acoustic branches dominate the $\kappa_L$ for all considered strain, about 96\%. Therefore, we only show contributions  from ZA, TA and LA acoustic branches vs strain in \autoref{mode-c}.
The contribution from ZA mode  increases significantly with strain increasing, from 27.9\% to 60.1\%, while the contribution from TA/LA mode decreases.
For small (large) strains [$<$ ($\geq$) 6\%], the TA (ZA) mode has largest contribution. When strain reaches  10\%, ZA mode
contributes  more than LA and TA modes to the total $\kappa_L$.

\begin{figure}
  \includegraphics[width=8cm]{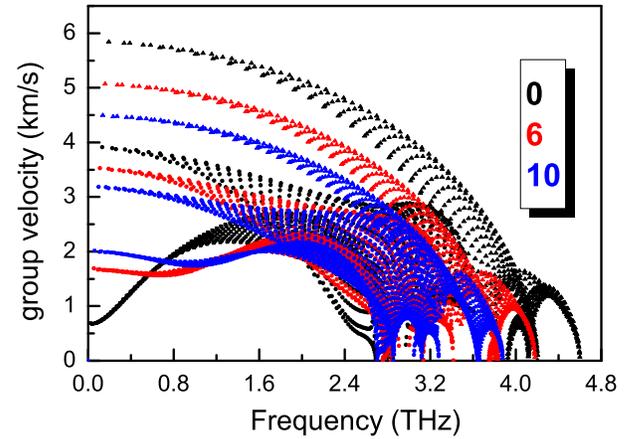}
  \caption{(Color online) The phonon mode group velocities of AsP monolayer  with strain   0\% (Black), 6\% (Red) and 10\% (Blue) in the first BZ for ZA (square symbol), TA (circle symbol) and LA (UpTriangle symbol) acoustic branches.}\label{v-c}
\end{figure}

To identify the underlying mechanism of strain-dependent phonon transport of AsP monolayer, phonon mode group velocities with  0\%, 6\% and 10\% strains   are plotted in \autoref{v-c}.
In low frequency region, group velocities  of ZA mode increase with increasing strain due to stiffened ZA dispersion. The   group velocity  of ZA mode  near $\Gamma$ point increases from   0.69  $\mathrm{km s^{-1}}$ to 1.70  $\mathrm{km s^{-1}}$ to 2.02  $\mathrm{km s^{-1}}$ with strain  from  0\% to 6\% to 10\%.
However, most of group velocities  of ZA mode decrease in high frequency region. For TA and LA branches, the reduction of phonon group velocities is observed due to softened phonon dispersions with increasing  strain. The largest  group velocity of TA (LA) mode reduces
from 3.91 (5.84)  $\mathrm{km s^{-1}}$    to 3.53 (5.07)  $\mathrm{km s^{-1}}$   to 3.19 (4.49) $\mathrm{km s^{-1}}$, when strain changes from 0\% to 6\% to 10\%.  The reduction of most phonon group velocities  would result in the decrease of $\kappa_L$ with increased strain.

\begin{figure*}
  \includegraphics[width=5.3cm]{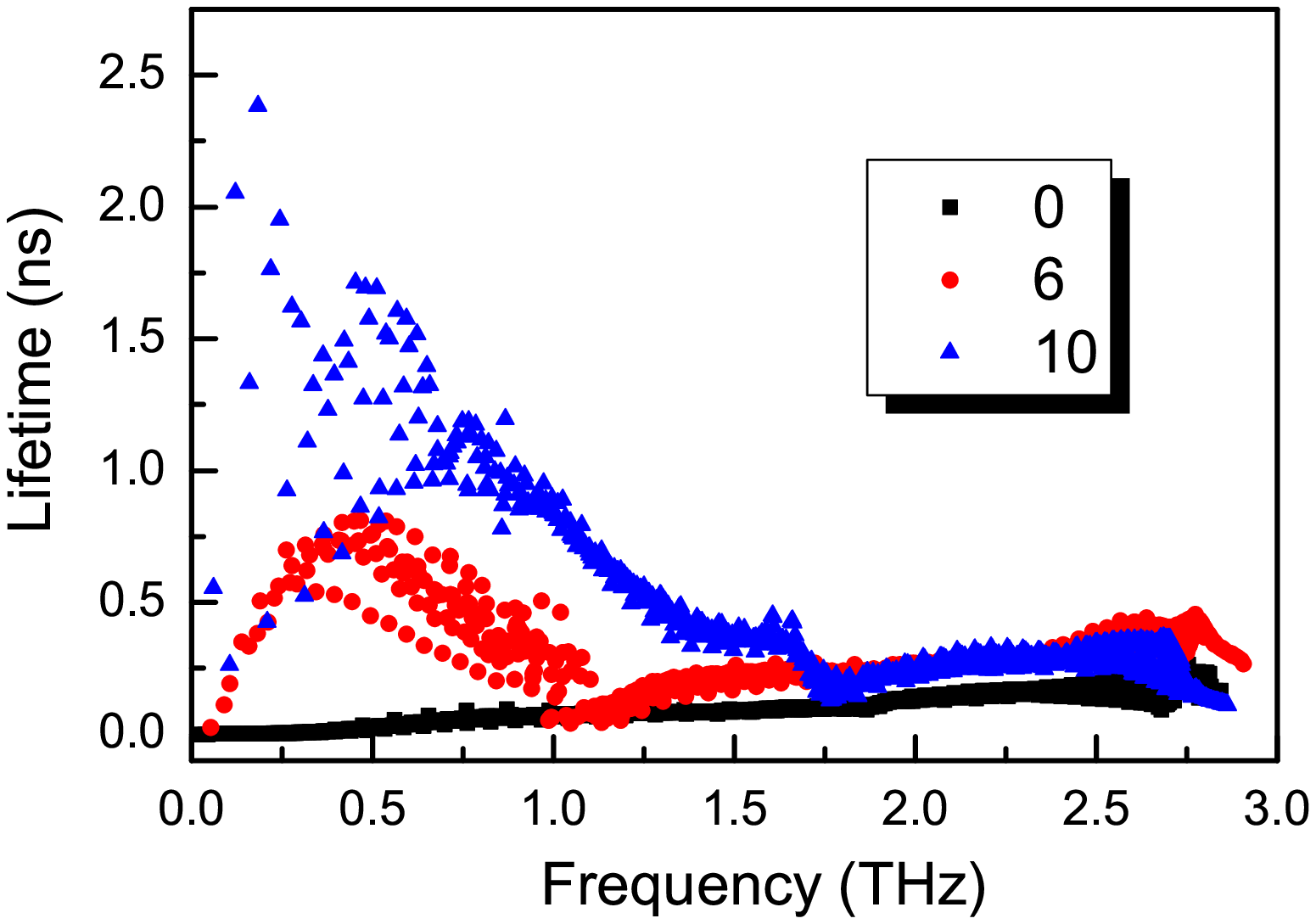}
    \includegraphics[width=5.3cm]{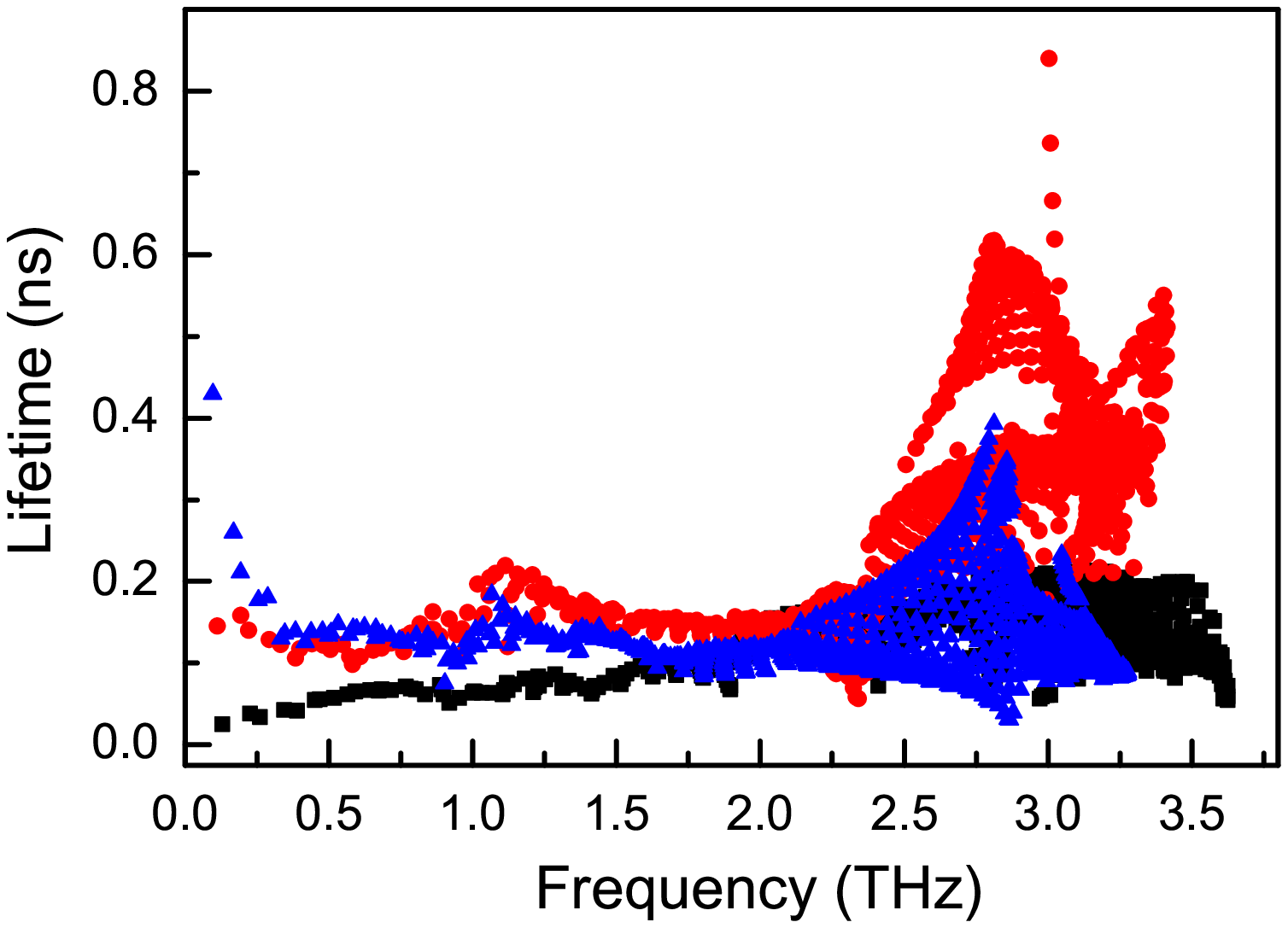}
      \includegraphics[width=5.3cm]{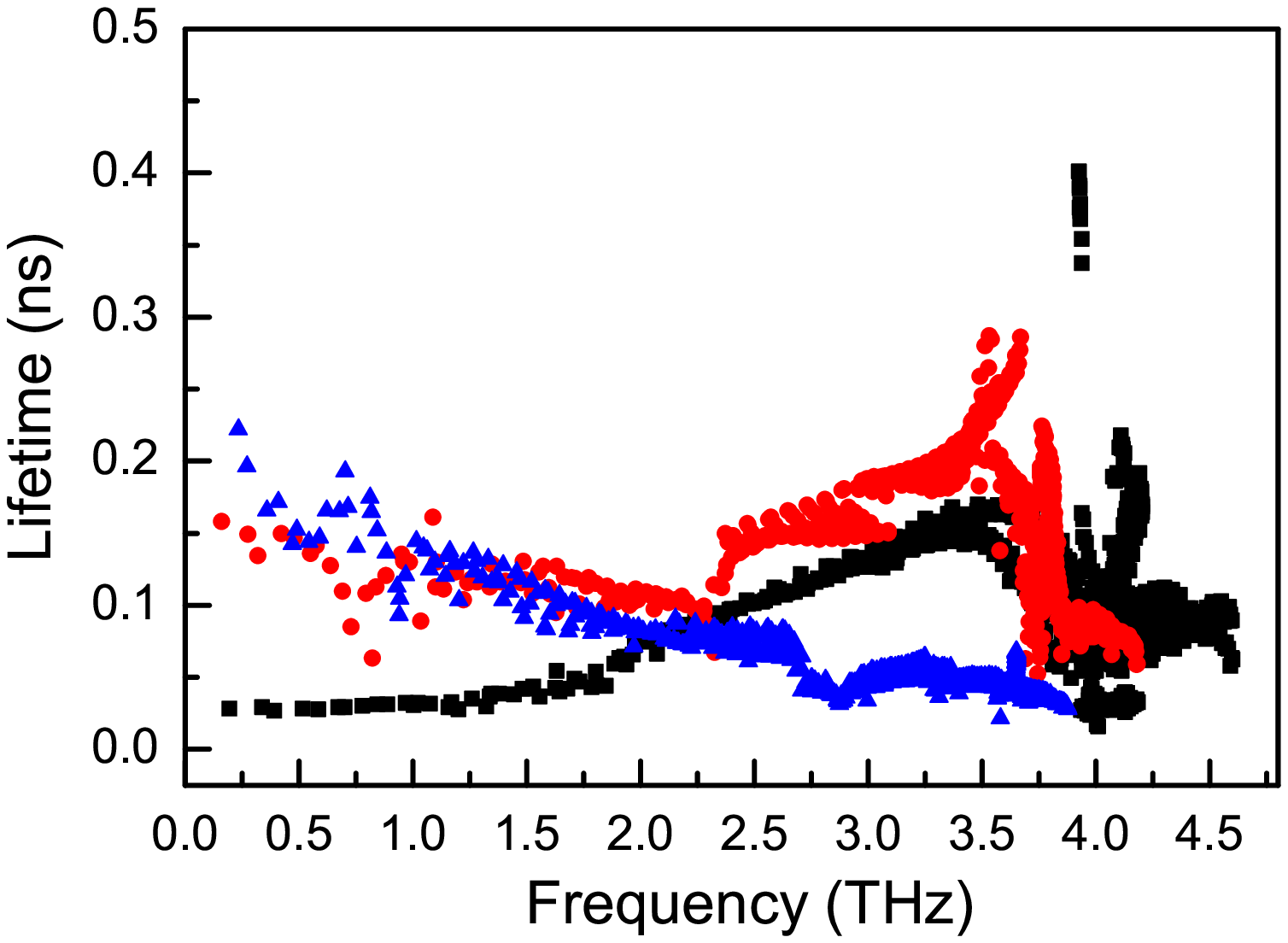}
  \caption{(Color online) The phonon mode lifetimes of AsP monolayer  with strain   0\% (Black), 6\% (Red) and 10\% (Blue) in the first BZ for ZA, TA  and LA acoustic branches from left to right.}\label{t-c}
\end{figure*}

Next, the  strain-dependent  phonon  lifetimes of AsP monolayer are calculated by  three-phonon scattering rate from  third-order anharmonic IFCs. The phonon lifetimes   with 0\%, 6\% and 10\% strains are plotted  in \autoref{t-c}. It is clearly seen that phonon  lifetimes of ZA mode present a monotonic increase  upon increased strain, which supports  the enhancement  of ZA contribution. However, the  phonon  lifetimes  firstly  increase for TA and LA branches, and then drop down.
By combining strain dependence of phonon group velocities
and phonon lifetimes, we can conclude that  at
low strains ($<$6\%) the  phonon lifetimes enhancement  is the major mechanism responsible for increased $\kappa_L$, while at high  strains
($>$6\%) the reduction of group velocities  as
well as the decrease of the phonon lifetimes of  LA and TA modes result in decreased $\kappa_L$.
At low strains ($<$6\%), the contribution from TA/LA has little change with increasing strain, which is because the enhanced phonon lifetimes partially neutralize  the negative effect of the group velocity reduction.  At high  strains ($>$6\%),  the contribution of  TA/LA rapidly decreases, which is due to the reduction of both  group velocities and  phonon lifetimes.

\begin{figure}
  \includegraphics[width=8.0cm]{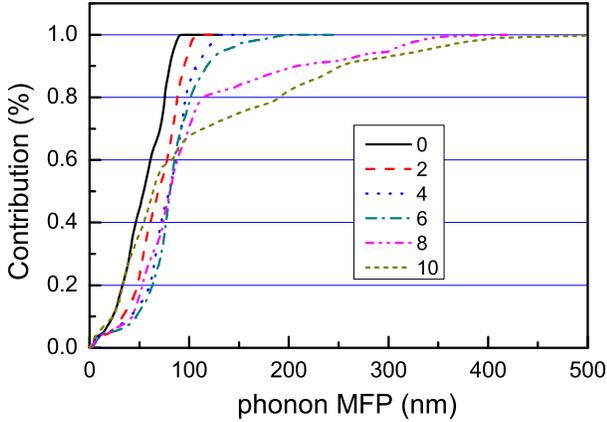}
  \caption{(Color online) From 0\% to 10\% strain, the cumulative lattice thermal conductivity divided by total lattice thermal conductivity with respect to phonon MFP at room temperature.}\label{mfp-c}
\end{figure}

To study size effect on $\kappa_L$, the cumulative $\kappa_L$ divided by total $\kappa_L$ with respect to MFP are  plotted in \autoref{mfp-c} with strain from 0\% to 10\% at 300 K.  It is well known that the cumulative $\kappa_L$ divided by total $\kappa_L$ increases, and then approaches one. It is clearly seen that
the saturated  MFP increases from 95 nm  to 490 nm  with strain from 0\% to 10\%, which means that the phonons with longer  MFP contribute to $\kappa_L$ with increased strain.  This demonstrates that strain can tune effectively    size effects on $\kappa_L$ of AsP monolayer.
The strain-tuned strong size effects on $\kappa_L$ can also be found in  antimonene, silicene, germanene and stanene\cite{l3,l11}..

\section{Conclusion}
In summary, the  first-principles calculations are performed to predict the strain-dependent $\kappa_L$ of  AsP monolayer.
The almost  the same $\kappa_L$ beween As and AsP monolayer is observed, which is because the reduce of phonon lifetimes from As to AsP monolayer  cancels out the group velocity enhancement. It is found that the increased tensile  strain  can harden  long wavelength ZA  acoustic branch, which may provide  guidance on
fabrication of AsP monolayer by tensile strain. In fact, the Bi monolayer  with the similar  graphene-like buckled structure has been  successfully  synthesized by tesile strain\cite{sci}.  Calculated results show that $\kappa_L$ of AsP monolayer firstly increases with strain increasing, and then decreases. The competition among  group velocity reduction,  phonon lifetime enhancement of ZA mode and nonmonotonic phonon lifetime of TA/LA mode leads to  up-and-down behavior of $\kappa_L$. Our work enriches  the studies of thermal transport of  2D materials with graphene-like buckled structure, and offers perspectives of tuning
thermal transport properties of AsP monolayer for applications such as thermoelectrics and nanoelectronics.

\begin{acknowledgments}
This work is supported by the National Natural Science Foundation of China (Grant No.11404391).  We are grateful to the Advanced Analysis and Computation Center of CUMT for the award of CPU hours to accomplish this work.
\end{acknowledgments}

\end{document}